\documentclass[twocolumn,prb,superscriptaddress,amsmath,amssymb,showpacs]{revtex4}

\usepackage{graphicx}
\usepackage{dcolumn}
\usepackage{bm}

\begin{document}


\title{Photon-spin qubit-conversion based on Overhauser shift of Zeeman energies in quantum dots}


\author{S. Muto}
\affiliation{CREST, Japan Science and Technology Agency, Kawaguchi 332-0012, Japan}
\affiliation{Faculty of Engineering, Hokkaido University, N13, W8 Kita-ku, Sapporo 060-8628, Japan}

\author{S. Adachi}
\affiliation{CREST, Japan Science and Technology Agency, Kawaguchi 332-0012, Japan}
\affiliation{Faculty of Engineering, Hokkaido University, N13, W8 Kita-ku, Sapporo 060-8628, Japan}

\author{T. Yokoi}
\affiliation{Faculty of Engineering, Hokkaido University, N13, W8 Kita-ku, Sapporo 060-8628, Japan}

\author{H. Sasakura}
\affiliation{CREST, Japan Science and Technology Agency, Kawaguchi 332-0012, Japan}

\author{I. Suemune}
\affiliation{CREST, Japan Science and Technology Agency, Kawaguchi 332-0012, Japan}
\affiliation{Research Institute of Electronic Science, Hokkaido University 
N21, W10 Kita-ku Sapporo, 001-0021 Japan}


\date{\today}

\begin{abstract}
We propose a new method to realize a conversion of photon qubit and spin qubit using the effective magnetic field created by the nuclear polarization known as Overhauser field. We discuss its preliminary experiment on InAlAs/AlGaAs self-assembled quantum dot and  also discuss effects of excitons which could destroy the conversion. 
\end{abstract}
\pacs{78.67.Hc, 71.70.Ej}

\maketitle
Recently, situation around quantum information processing has been greatly changed. 
Especially for quantum cryptography, node-to-node telecommunication over 100 km is now possible\cite{Kosaka03}. One ultimate goal of the quantum information processing is the quantum computing\cite{Feynman85,Deutsch85}. Although the proposed application of quantum computer (QC) is still restricted to some special uses such as prime factorization\cite{PWShor94} and database search\cite{Grover97}, new attractive ones will come out when the practical hardware is realized.
 One of the candidates for a qubit of the quantum computation is the electron spin\cite{Loss98,Sasakura01} (e-spin). Therefore, it is desirable to convert an electron-spin qubit to a photon qubit. An immediate application is to realize the quantum repeater to multiply the distance of quantum telecommunication. Besides, if a large scale computation is realized using electron spins alone, then we can connect the e-spin QC to quantum network of photons. It is also possible that electron-qubits work as an intermediator between photon qubits and nuclear-spin qubits. This could be an important application since the nuclear spin qubits with their ultra-long coherence is expected to work as memory qubits for any kind of QCs.

\begin{figure}
\includegraphics[width=250pt]{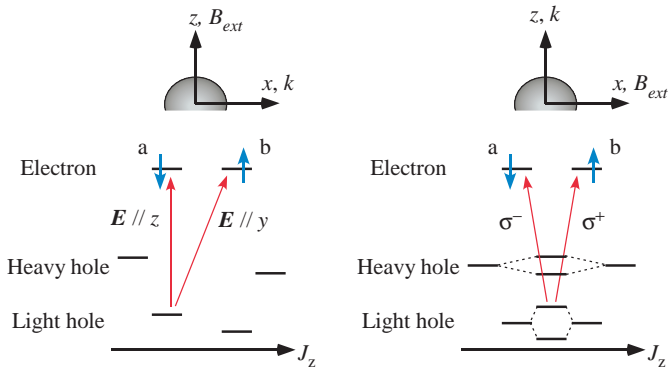}
\caption{Photon-spin qubit conversion using optical selection rule of electron-light hole transition. Electron spin states are degenerate due either to zero electron $g$-factor (refs.~\onlinecite{Vrijen01} and \onlinecite{Kosaka01}) or to the nuclear field (this study). i) Applied field is in the growth ($z$-) direction of a self-assembled quantum dot and photon is incident in $x$-direction. ii) Photon is incident in the $z$ direction and the applied field is in the $x$ direction.}
\label{Figure1}
\end{figure}

Yablonovitch and coworkers\cite{Vrijen01} have already proposed the qubit conversion between electron spins and photons based on the selection rules of optical transitions in quantum wells and quantum dots using the "$g$-factor engineering\cite{Kosaka01}". They also proposed a quantum repeater\cite{Yablonovitch03} based on this idea and using their e-spin QC\cite{Vrijen00}. Their idea of $g$-factor engineering is to chose proper semiconductor materials or their combinations to realize "zero $g$-factor" for electrons. By applying a static magnetic field, we can realize degenerate electron-spin states and non-degenerate hole states (see Fig.~\ref{Figure1}). Therefore, we can transform a photon qubit to an electron-spin qubit multiplied by one of split hole states. 
\begin{equation}
a\left| 0 \right\rangle  + b\left| 1 \right\rangle  \Leftrightarrow \left( {a\left|  \downarrow  \right\rangle  + b\left|  \uparrow  \right\rangle } \right) \otimes \left| {\rm hole} \right\rangle \nonumber
\end{equation}
Here, $\left| 0 \right\rangle$ and $\left| 1 \right\rangle$ are two basis of a photon qubit and $\left|  \downarrow  \right\rangle$, $\left|  \uparrow  \right\rangle$ are up and down states of an e-spin qubit.

One topic related to this is the electrical control of the $g$-factor using parabolic quantum well\cite{Gossard99}. Kato and coworkers\cite{Kato03} showed that zero $g$-factor is actually realized for electrons by applying a proper voltage to the quantum well.
 Those ideas, though elegant, need special choices of semiconductor material and structures. It will be beneficial to have a more general way regardless of materials to make electron $g$-factor "effectively zero". Here we propose a new way using Overhauser shift\cite{Overhauser53} of Zeeman energies known also as the "dynamic nuclear polarization". For this, all we need is the circularly polarized light and static magnetic field.\\

\textbf{Basic idea}: Here we consider a In(Al)GaAs/GaAs quantum dot as an example. The basic idea is that the electrons feel both the external field and the nuclear field, that is the effective magnetic field caused by the nuclear spin polarization through the hyperfine interaction with electron spins, while the holes feel only the external field. Consequently, if we tune either the external field or the nuclear field such that they cancel each other, an electron feels no magnetic field and a hole feels a non-zero field. Hence the situation of zero $g$-factor for electrons is effectively realized. Later, we will see that the nuclear field to holes does not have to be zero. It is sufficient that the nuclear fields for both carriers is different.\\

\textbf{Details of proposal}: Hamiltonian for the electron-hole pairs including nuclear spins is written\cite{Abragam61,Bayer02} to the lowest order of electron and hole spins by,
\begin{eqnarray}
{\cal H} &=& \bm s_e  \cdot \mu_B  \tensor{g}_e  \cdot \bm B + \bm j_h  \cdot 2 \tensor{\kappa} \bm B + {\cal H}_{HF} \nonumber\\
&+& \rm{(spin-independent \ term)}
\label{eq1}
\end{eqnarray}
Here, $\tensor{g}_e$ and $\tensor{\kappa}$ are diagonal tensors corresponding to Zeeman energies for an electron and a hole, and hyperfine interaction   between electron spin and nuclear spin is written as 
\begin{equation}
{\cal H}_{HF}  = \sum\limits_n { - \left( {\frac{{\mu_I }}{I}} \right)_n } \bm H_e^{(n)}  \cdot \bm I_n 
\label{eq2}
\end{equation}
using nuclear magnetic moment $\mu_I$ and the effective field for the $n$-th nuclear spin $\bm I_n$,
\begin{equation}
\bm H_e^{(n)}  =  - 2\mu_B \left\{ {\frac{1}{{r_n ^3 }}\left( {\bm l - \bm s} \right) + \frac{{3\bm r_n \left( {\bm s \cdot \bm r_n } \right)}}{{r_n^5 }} + \frac{{8\pi }}{3}\bm s \delta \left( {\bm r_n } \right)} \right\},
\label{eq3}
\end{equation}
where $\bm r_n = \bm r - \bm R_n$ is the position vector of electron (hole) measured from the $n$-th nucleus and $\bm l$, $\bm s$ are corresponding angular momentum and spin of an electron (a hole).
Remembering that both electron and hole (heavy or light) have definite angular momentum $l$ (0 for electron and 1 for hole), $s$ (=1/2), $j$ (=3/2) and $j_z$ ($\pm$3/2 for heavy holes and $\pm$1/2 for light holes), we can project the effective field to the direction of $\bm j$ ($\bm s$ for electrons). Namely, 
\begin{equation}
-\frac{\mu_I}{I} \bm H^{(n)}_e= a_{j}^{(n)} \bm j
\label{eq4}
\end{equation}

According to Abragham\cite{Abragam61}, 
\begin{equation}
a_{j} ^{(n)}  = \frac{{16\pi }}{{3I}}\mu _{B} \mu _{I} \left| {\psi \left( {\bm R_n } \right)} \right|^2  + \frac{{2\mu _{B} \mu _{I} }}{I}\left\langle {\frac{1}{{r_n ^3 }}} \right\rangle \frac{{l\left( {l + 1} \right)}}{{j\left( {j + 1} \right)}}
\label{eq5}
\end{equation}
Here, $\left\langle {\frac{1}{{r_n ^3 }}} \right\rangle$ is the average over electron (hole) wave function. Note that the second term is 0 for electrons and that the first term is 0 for holes. Therefore,
\begin{eqnarray}
{\cal H}_{HF}  &=& \sum\limits_n {a_{j} ^{(n)} } \bm j \cdot \bm I_n    \label{eq6}\\ 
  &=& \bm s_e  \cdot \sum\limits_n {\frac{{16\pi }}{{3I}}} \mu_{B} \mu_{I} \left| {\psi \left( {\bm R_n } \right)} \right|^2 \bm I_n  \nonumber\\
  &+& \bm j_h  \cdot \sum\limits_n {\frac{{16}}{{15I}}} \mu_{B} \mu_{I} \left\langle {\frac{1}{{r_n ^3 }}} \right\rangle \bm I_n 
  \label{eq7}
\end{eqnarray}
Here, $\bm s_e$ and $\bm j_h$ indicate the electron spin and the total angular momentum of a hole. After all, the spin-dependent terms of Hamiltonian is 
\begin{eqnarray}
{\cal H} &=& \bm s_e  \cdot \left\{ {\mu_B \tensor{g}_e  \cdot \bm B + \sum\limits_n {\frac{{16\pi }}{{3I}}} \mu_{B} \mu_{I} \left| {\psi \left( {\bm R_n } \right)} \right|^2 \bm I_n } \right\} \nonumber\\
&+& \bm j_h  \cdot \left\{ {2\tensor{\kappa} \bm B + \sum\limits_n {\frac{{16}}{{15I}}\mu_{B} \mu_{I} \left\langle {\frac{1}{{r_n ^3 }}} \right\rangle \bm I_n } } \right\}
  \label{eq8}
\end{eqnarray}

The effective field to electron can be written in a familiar form of contact term as
\begin{eqnarray}
&&\sum\limits_n {\frac{{16\pi }}{{3I}}} \mu_{B} \mu_{I} \left| {\psi \left( {\bm R_n } \right)} \right|^2 \left\langle {\bm I_n } \right\rangle  = \sum\limits_i {A_i } \bm I_{ave}^i \label{eq9}\\
&&A_i  = \frac{{16\pi }}{{3I}}\frac{{\mu _{B} \mu _{I} }}{{v_o }}\eta _i ,\ \eta_i  = \left| {u(\bm R_i )} \right|^2 \label{eq10}\\
&&\psi \left( \bm R \right) = \Psi \left(\bm R \right)u\left( \bm R \right)\label{eq11}\\
&&\bm I_{ave}^i  =\frac{ \int {d^3 } \bm R \left\langle {\bm I^i \left( \bm R \right)} \right\rangle \left| {\Psi \left( \bm R \right)} \right|^2} {\int {d^3 } \bm R\left| {\Psi \left( \bm R \right)} \right|^2} \label{eq12}
\end{eqnarray}
where $\left\langle {\bm I_n } \right\rangle$ indicates the quantum mechanical average of nuclear spin, and    and $\Psi \left( \bm R \right)$ are the envelope function and Bloch amplitude of an electron. Here, $i$ indicates the nuclear site in a unit cell with volume $v_0$ and the sum runs over all the nuclear site of a unit cell.

If we apply an external field only in the $\nu$ ($\nu= x \ {\rm or} \ z$) direction, then $\left\langle {\bm I_n } \right\rangle$ and $\bm I_{ave}^i$ are also in the $\nu$-direction. Therefore, electrons feel the effective filed $B_{eff,e}$,
\begin{equation}
g_{e,\nu } \mu _B B_{eff,e}  = g_{e,\nu } \mu_B B_{ext}  + \sum\limits_i {A_i I_{ave}^i } 
\label{eq13}
\end{equation}
On the other hand, holes feel the effective field, $B_{eff,h}$,
\begin{equation}
2\kappa _\nu  B_{eff,h}  = 2\kappa _\nu  B_{ext}  + \sum\limits_n {\frac{{16}}{{15I}}} \mu_{B} \mu_{I} \left\langle {\frac{1}{{r_n ^3 }}} \right\rangle \left\langle {I_n } \right\rangle 
\label{eq14}
\end{equation}
Therefore, if we can tune either $B_{ext}$ or $I_{ave}^i$, so that 
\begin{equation}
g_{e,\nu } \mu _B B_{ext}  + \sum\limits_i {A_i I_{ave}^i }  = 0,
\label{eq15}
\end{equation}
electrons feel zero magnetic field.

Then the splitting for holes given by eq.~\ref{eq1} is, using eq.~\ref{eq15} in eq.~\ref{eq14} and multiplying by 
$\displaystyle 2\left| {j_z } \right|$, 

\begin{eqnarray}
4\left| {j_z } \right|\kappa _\nu  B_{eff,h}  &=&  - \left( {\frac{{4\left| {j_z } \right|\kappa _\nu  }}{{\mu _B g_{e,\nu } }}} \right) \cdot \sum\limits_i {A_i I_{ave}^i } \nonumber \\
 &+& \sum\limits_n {\frac{{32\left| {j_z } \right|}}{{15I}}} \mu _B \mu _I \left\langle {\frac{1}{{r_n ^3 }}} \right\rangle \left\langle {I_n } \right\rangle  \nonumber\\
 &\sim&  - \left( {\frac{{4\left| {j_z } \right|\kappa _\nu  }}{{\mu _B g_{e,\nu } }}} \right)\sum\limits_i {A_i I_{ave}^i } 
\label{eq16}
\end{eqnarray}
for both heavy ($\left| {j_z } \right|= 3/2$) and light ($\left| {j_z } \right|=1/2$) holes. Since the dipolar terms is usually much smaller than the contact term, the second term in the middle of eq.~\ref{eq16} can be neglected. However, we note that this term does not have to be negligible for our proposal. 

\begin{figure}
\includegraphics[width=200pt]{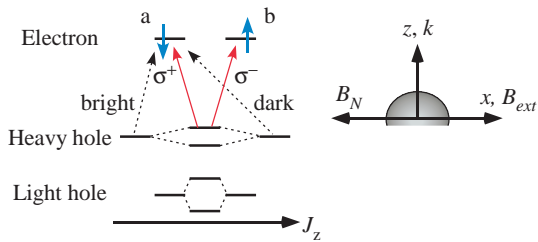}
 \caption{Photon-spin qubit conversion using optical selection rule of electron-heavy hole transition and nuclear field . Photon is incident in the $z$-direction and the applied field is in the $x$-direction.
Here, $\displaystyle B_N  = \sum\limits_i {A_i I_{ave}^i} /g_e \mu_B $
  indicates the nuclear field to an electron.
Dotted arrows indicate bright and dark excitons for the for the $\left|  \downarrow  \right\rangle $ electron.}
\label{Figure2}
\end{figure}

The resultant band diagram is just as shown by Fig.~\ref{Figure1} (ref. \onlinecite{Kosaka03}) and Fig.~\ref{Figure2}. In Fig.~\ref{Figure1}(a), the external field was applied to the $z$ (growth) direction and photons are introduced in the $x$ (layer) direction. In Fig.~\ref{Figure1}(b), and Fig.~\ref{Figure2}, the directions are exchanged. Here we have neglected an important effect of exciton which will be discussed later.\\

\textbf{Experiments}: Recently, we have observed Overhauser shift $\sum\limits_i {A_i I_{ave}^i }$ in InAlAs QD grown by Stranski-Krastanow mode of MBE\cite{Yokoi04}. The experiment demonstrated Overhauser energy of 19 $\mu$eV which is controllable by the excitation polarization and more generally by "optical orientation\cite{OptOrientation}". We have also observed for the structure that the heavy hole $g$-factor in the $x$-direction is nearly equal to the electron $g$-factor (more precisely, $\left| {g_{hh,x} } \right| \cong \left| {g_{e,x} } \right| \cong 0.43$). Here, $\mu_B g_{hh,x}$ is $4\left| {j_z} \right|\kappa_x  \left( { = 6\kappa_x } \right)$. Equation~\ref{eq15} indicates that we can cancel the Overhauser field for electrons with $B_{ext}$ = 0.74 T, and that we can create the energy splitting for holes as much as Overhauser shift, which is about 19 $\mu$eV in Fig.~\ref{Figure2}. Although this is not a large energy, we can distinguish two hole levels with this separation since the intrinsic energy width of excitons are reported to be as small as 4 $\mu$eV\cite{Birkedal01}. Here we note that the $g_{hh,x}$ is non-zero. This is quite in contrast to the quantum wells where the in-plane $g$-factor for heavy hole is zero\cite{Vrijen01}. However, Bayer and coworkers also observed non-zero $g_{hh,x}$ of $-0.35$ ($g_{e,x}$ = $-0.65$) for their InGaAs self-assembled quantum dot\cite{Bayer00}.\\

\textbf{Exciton effects}: Here we discuss the excitonic effect which is inevitable in the current structure. Unfortunately, the structure cannot be used for the qubit-conversion due to the electron-hole exchange interaction of excitons. In the heavy-hole exciton formed in Fig.~\ref{Figure2}, the exchange interaction splits of exciton states (with energy $\delta_0$) to the dark and bright excitons, corresponding to parallel and anti-parallel alignment of electron and hole spins ($\bm s_e$ and $\bm j_h$), in the absence of magnetic field. This indicates that a photon with combination of right-circular polarization and left-circular polarization is converted to combination of two bright excitons, 
$a | s_z =-1/2>|j_z =3/2> + b|s_z=1/2>|j_z =-3/2>$  which is an entangled state of electron-hole pairs, and obviously neither a spin qubit of an electron nor of a hole can be factored out. Even in the presence of magnetic field, this situation persists as long as $g_h \mu_B B_{ext} < \delta_0$. In our measurements\cite{Yokoi04}, $\delta_0$= 39 $\mu$eV and $g_h \mu_B B_{ext}$ = 19 $\mu$eV, and therefore, we are still in a region where bright and dark exciton persists and the qubit conversion does not work. Therefore, we should either increase the nuclear polarization (currently on the order of 10 \%) and resultant Overhauser shift or reduce the exchange energy. Apart from the self-assembled QDs, Gammon and coworkers reported\cite{Gammon01} the Overhauser shift of 90 $\mu$eV (nuclear polarization of 65 \%) for the GaAs QD using the well width fluctuation of a GaAs/AlGaAs quantum well. Also, we can reduce the exchange interaction by spatially separating electron-hole pairs. The spatial separation will also help to reduce the adverse effects of possible anisotropic exchange interaction which couples the up and down spin states of an electron\cite{Bayer02}. However, as the separation reduces the probability of optical transitions, optical cavity to enhance the optical transitions will be necessary. 

In summary, we proposed a new method to realize conversions between a spin qubit and a photon qubit. The method uses Overhauser field by the nuclear polarization which is controlled by the external optical excitation, and is regarded to be more flexible than the previous proposal.

The authors are grateful to Prof. K. Yoh of Hokkaido University for stimulating discussions. S. M. also thanks Ms. M. Konno for typing of mathematical formula. This work was partially supported by a Grant-in-Aid for Scientific Research in Priority Areas "Semiconductor Nanospintronics" of The Ministry of Education, Culture, Sports, Science and Technology, Japan.

\end{document}